\documentclass[aps,pra,twocolumn,superscriptaddress,groupedaddress]{revtex4}  
\usepackage{graphicx}  
\usepackage{dcolumn}   
\usepackage{bm}        
\usepackage{amssymb}   
\usepackage{amsmath}
\usepackage{dsfont}
\usepackage{cancel}
\usepackage{empheq}
\usepackage{graphicx,color}
\usepackage{accents}
\usepackage{wrapfig}
\usepackage[caption=false]{subfig}
\usepackage{hyperref}
\makeatletter
\renewcommand*\env@matrix[1][*\c@MaxMatrixCols c]{%
  \hskip -\arraycolsep
  \let\@ifnextchar\new@ifnextchar
  \array{#1}}
\makeatother

\begin{document}

\title{A Physical Perspective on Classical Cloning}
\author{Anirudh Reddy$^{1}$, Joseph Samuel$^{1}$, Supurna Sinha${}$}
\address{Raman Research Institute, Bangalore 560080, India.
}
\date{\today}

\begin{abstract}
The celebrated quantum no-cloning theorem states that an arbitrary quantum state
cannot be cloned perfectly. This raises questions about cloning of classical states, which have also
attracted attention.
Here, we present a physical approach to the classical cloning process showing how cloning can be realised
using Hamiltonians. After writing down a canonical transformation
that clones classical states, we show how this can be implemented by Hamiltonian evolution.
We then propose an experiment using the tools of nonlinear optics to realise the ideas
presented here. Finally, to  understand the cloning process in a more realistic context,
we introduce statistical mechanical noise to the system and study
how this affects the cloning process.
While most of our work deals with linear systems and harmonic oscillators,
we give some examples of cloning maps on manifolds and show that any system whose configuration space
is a group manifold admits a cloning canonical transformation.

\end{abstract}

\maketitle

\section{Introduction}

Wootters and Zurek's \cite{wootters} work on the no-cloning theorem has led to extensive research 
on the quantum cloning process and its physical implications. While the studies in the 
quantum regime are both abstract 
\cite{hillery, buzek, gisin, werner, RevModPhys.77.1225} and 
application-based \cite{kempe, simon, fasel, lamas, agarwal}, work on 
the classical cloning process 
has been extremely limited and restricted to a purely mathematical approach \cite{plastino, aaron, nich, matsumoto}. 
There appears to be a belief that the classical cloning process is trivial, 
perhaps because it is so familiar. Computers routinely copy files, photocopying machines are widespread
and the genetic information contained in DNA is replicated every time a cell divides.
However, there are subtleties  related to classical cloning\cite{aaron} 
and even a classical statistical no-cloning theorem\cite{plastino} proved under certain, 
general assumptions. A good understanding of
the copying of classical information is essential 
to appreciate the quantum case and the relation between the two.

The discussion of cloning involves three coupled systems: a source, a target and a machine. The source contains the state
to be cloned.  The target is initially in a standard blank state and the machine is initially in a standard ready state, 
both independent of the source. The objective 
of cloning is to copy the state of the source into the target, without destroying the original. In the copying process, the
machine state may be altered and has to be reset before the next copy can be made. In this paper, we consider the cloning process
from a physical point of view, 
clarifying the conditions under which classical cloning is 
possible and explicitly constructing
Hamiltonians which implement the cloning process.

Before going any further we need to define more precisely 
what we mean by a ``state''. A ``state'' in classical mechanics is defined as a point
in phase space. A system in classical mechanics has a configuration space $Q$ with local coordinates $q^r$ 
and the phase space has twice the dimension $(q^r,p_r)$ including  coordinates as well as conjugate momenta.
Phase spaces are even dimensional and allow us to define Poisson brackets between functions. 
In statistical physics, a state would be defined as a probability distribution on phase space.
Daffertshofer et al. \cite{plastino} have proved a classical no-cloning theorem when states are regarded as  
probability distributions. This proof is based on the invariance 
of the relative entropy (Kullback-Leibler divergence) under arbitrary diffeomorphisms of the phase space.
It follows from their work that cloning of classical states is forbidden whenever the relative entropy of the total system is
well defined. However, there are situations where the relative entropy of the system is ill defined, for example when the phase space
distributions have delta function support and this permits a discussion of classical cloning.

Aaron Fenyes \cite{aaron} studied the cloning process from the viewpoint of symplectic geometry. In his treatment, 
classical states are points in a symplectic manifold and the cloning process is a symplectomorphism.
This provides a very general setting for the cloning process in classical mechanics. 
In the physics community, symplectic manifolds and symplectomorphisms are more usually referred to as phase spaces and canonical transformations.
In this paper, we use the framework provided by Fenyes to study the classical cloning process 
in more detail. This paper is organised as follows. In Sec. II we summarise the 
classical cloning process using a symplectic map.  We describe a procedure for generating a 
symplectic cloning map for the case of variables in a linear phase space. 
We show how to systematically generate Hamiltonians 
to realise a given linear symplectic map.
We then prove a general result that the phase spaces normally occurring
in classical mechanics (those that emerge from a configuration space) admit a cloning map.
In Section III, we propose an experimental realisation of a cloning machine using nonlinear optics.
We then explore deviations from  this ideal situation and introduce statistical mechanical noise and look at the 
effect of temperature on the cloning process in Sec IV. The introduction of statistical mechanical noise does result 
in a corruption of the cloning process, consistent with the no-cloning theorem in the classical statistical 
domain as expected from Ref. \cite{plastino}. 
Finally, in Sec V we end the paper with concluding remarks.

\section{Classical Cloning by Canonical Transformations}

\textbf{Definition of the classical cloning process}: Following \cite{aaron}, let $M$ and $N$ 
be symplectic manifolds. (A symplectic manifold is a manifold with a closed, non degenerate two-form $\omega_0$. 
Diffeomorphisms that preserve the two form $\omega_0$ are called symplectomorphisms.) Let $(N,N,M)$ represent the source, target and machine
respectively. Initially, we suppose that the target and machine are in 
standard states $b$ and $r$. Given an arbitrary state $s\in N$ of the source, 
a cloning map is a symplectomorphism 
$\psi:N \times N \times M \to N \times N \times M$ such that $\psi(s,b,r)=(s,s,f(s,b,r))$ 
for all $s \in N$, where $b,r$ are independent of $s$ \cite{aaron}. 
Here the manifold $M$ acts as the copying machine, while the source and the target states are on the manifold $N$. 
The source state is $s$, the material to be copied (for instance a birth certificate!); 
$b$ the target state, which is initially blank, as are the A4 sheets 
in the tray of the copying machine;  and the machine state is $r$  ($r$ for ready) before cloning. 
We would like to know whether there exists a cloning map for a given classical system $(N,\omega_0)$. What choice of the machine $M$ is needed to achieve this?
It is also of interest to determine how these maps can be generated in the laboratory by physical processes.

Let us suppose that that there is a cloning map $\psi$ as above.
Let us now fix $s=s_0$ and consider the linearised map $\phi$ that maps the tangent space of $(s_0,b,r)$ to the 
tangent space of $(s_0,s_0,f(s_0,b,r))$. These tangent spaces are symplectic vector spaces and
$\phi$ is a linear symplectic cloning map. Thus, the existence of $\psi$ would imply the existence of 
linear symplectic cloning maps.
Let us begin by addressing the simpler problem of linear symplectic cloning.
Linearity results in a considerable simplification of the problem and permits explicit construction of cloning maps. 
As we will see later, this simple case illuminates the more general problem of classical cloning. 
It also covers the physically important case of harmonic
oscillators, which are easily realised in an optics laboratory as modes of the electromagnetic field.

Let us start with the simplest example and  
choose $M$ and $N$ to be two dimensional symplectic 
vector spaces $(\mathds{R}^2,\omega_0)$, 
so that we can view $M$ and $N$ as phase spaces, 
with each point in these spaces being 
labelled by a position and a momentum. 
Let $b=r=\begin{pmatrix} 0 \\ 0 \end{pmatrix}$, and 
$s=\begin{pmatrix} q^s \\ p_s \end{pmatrix}$. A linear symplectic cloning map on $\mathds{R}^6$ is given by

\begin{equation}
\phi(s,b,r)=\begin{bmatrix}[cc|cc|cc]
 1 & 0 & 1 & 1 & -1 & 1 \\
 0 & 1 & 1 & 2 & 0 & 1 \\ \hline
 1 & 0 & 0 & 1 & 0 & 1 \\
 0 & 1 & -1 & -1 & 1 & 0 \\ \hline
 1 & 0 & 1 & 2 & -1 & 2 \\
 0 & -1 & 0 & -1 & -1 & -1 
\label{clonmatrix}
\end{bmatrix}
\renewcommand\arraystretch{2}
\begin{bmatrix}
s\\ \hline
b\\ \hline
r\\
\end{bmatrix}.
\end{equation}

It is a cloning map because it satisfies $\phi(s,b,r)=(s,s,Fs)$, where $F=\begin{bmatrix} 1 & 0 \\ 0 & -1 \end{bmatrix}$. $\phi$ is a symplectic map as it satisfies the condition 
$\phi^T\Omega\phi=\Omega$, where 
$\Omega=dq^s\wedge dp_s+dq^t\wedge dp_t+dq^m\wedge dp_m$ 
is the symplectic form on $\mathds{R}^6$. 
Cloning by a machine is only possible if the dimension of $M$ is  
greater than or equal to the dimension of $N$.  A minimal choice is $M=N$. 
As emphasised by \cite{aaron}, cloning is impossible without the presence
of the machine.

\subsection{Cloning as a Canonical Transformation}
In this section, we discuss a systematic procedure 
for the generation of cloning maps on the symplectic vector space $\mathds{R}^2\times\mathds{R}^2\times\mathds{R}^2$.
By definition, the cloning map $\phi:\mathds{R}^6\to\mathds{R}^6$ must send $(s,b,r)\to(s,s,Fs)$ 
(where, $b=0$ and $r=0$ are at the origin). 
The two dimensional  vector subspace  $V$ of $\mathds{R}^6$ spanned by vectors of the form $(s,0,0)$ is
mapped to the two dimensional subspace $W$ of $\mathds{R}^6$ spanned by vectors of the form $(s,s,Fs)$. 
In order for the map from $V$ to $W$ to preserve the 
symplectic form,
$F$ must be antisymplectic, i.e, 
it must reverse the symplectic structure. A simple choice for $F$ is  
$F=\begin{bmatrix} 1 & 0 \\ 0 & -1 \end{bmatrix}$. 
We need to now extend this map to all of $\mathds{R}^6$. Clearly $V^c$, the 
symplectic complement of $V$ must map to $W^c$, the symplectic complement of $W$. A systematic procedure for constructing the map is the  
Gram-Schmidt procedure \cite{mcduff1998introduction}. This procedure is carried out in detail in Appendix B to produce the cloning map 
above (\ref{clonmatrix}).

Since the Gram-Schmidt procedure involves choices there is clearly ambiguity in the extension of the cloning map. What is the extent of this ambiguity?
We have the freedom to apply any symplectic transformation to the combined target and machine
state before cloning. As a result
there is an $Sp(4)$ worth of ambiguity in mapping $V^c$ to $W^c$. In addition, we also have the freedom to compose $F$ with any other symplectic transformation in
$\mathds{R}^2$ of the machine. 
Thus there is a total of $Sp(4)\times Sp(2)$ worth of cloning maps in $\mathds{R}^2$.

\subsection{Hamiltonian Cloning}
Having found a linear cloning map, we would like to implement this transformation by 
a Hamiltonian, so that cloning can be realised in a laboratory. Since we are working with linear spaces, 
it is natural to consider quadratic Hamiltonian functions. If $x$ is a vector in $\mathds{R}^6$, ($x^i, i=1,6)$, our 
Hamiltonian is a quadratic function 
\begin{equation}
H(x)=1/2x^i h_{ij} x^j 
\label{hfunction}
\end{equation}
with $h_{ij}$ a real symmetric
matrix $h_{ij}=h_{ji}$. Using Hamilton's equations we get an evolution
\begin{equation}
\dot{x}^i= \Omega^{ij} h_{jk} x^k
\label{hamilton}
\end{equation}  
which is a linear transformation generated by $h^i{}_k=\Omega^{ij} h_{jk}$. 
Under time evolution for a time $t$, the vector $x$ would be mapped
to the vector $[\exp{h t}]x$ where $h$ is the matrix $h^i{}_k$.

We will now explicitly construct 
Hamiltonians to implement the map $\phi$ mentioned earlier.
The map $\phi$ cannot be 
realised via a single time independent quadratic
Hamiltonian. 
\footnote{Mathematically, this is because the exponential map from
the Lie algebra of the symplectic group to the group is not 
surjective \cite{hilgert}.}. 
To see this, suppose that the symplectic map $\phi$ and
its corresponding Hamiltonian matrix $h$ are related by
\begin{equation}
\phi=e^{ht}.
\end{equation}
Taking the logarithm of both sides, 
we find that $\log{\phi}=ht$, where the logarithm is the matrix log. 
We find that the left hand side is complex, while the Hamiltonian matrix must be real. 
This contradiction shows that to realise the cloning map mentioned above we need at least 
two Hamiltonians. In fact,
Using the polar decomposition of symplectic matrices \cite{hilgert},
we can write $\phi= \exp{X}\exp{Y}$ where $X,Y$ are in the Lie algebra of the symplectic group.
Writing $h_1=X/\tau$ and $h_2=Y/\tau$, we can express the cloning map as

\begin{equation}
\phi=e^{h_1\tau}e^{h_2\tau},
\end{equation}
where $\tau$ will be chosen later to suit our convenience.
The Hamiltonian matrices in explicit numerical form 
(rounded to three decimal places) are ,

\begin{equation}\label{ham1}
h_1\tau =
\begin{bmatrix}
 -0.209& -0.003 & -0.206 & -0.332 & 0.206 & -0.128 \\
 0.418& 0.209 & -0.120 & -0.120& -0.006 & 0.006 \\
 0.120 & -0.332 & -0.738 & -0.254& 0.284 & -0.583 \\
 -0.120 & 0.206 & 1.066& 0.738 & -0.535 & 0.738 \\
 -0.006 & -0.128 & -0.738& -0.583& 0.409 & -0.505\\
 -0.006 & -0.206 & -0.535 & -0.284 & 0.254 & -0.409 \\
\end{bmatrix}
\end{equation}

\begin{equation}\label{ham3}
h_2\tau=
\begin{bmatrix}
 0.779 & -0.203 & -0.796 & 0.834 & 0.117 & 0.329 \\
 0.101 & -0.779 & -1.438& -0.412 & 1.107 & -1.741 \\
 0.412 & 0.834 & -2.509 & -0.722 & 2.479 & 0.563 \\
 -1.438& 0.796 & 4.039 & 2.509 & -1.512 & 3.013 \\
 1.741 & 0.329 & -3.013& 0.563 & 1.534 & 0.958 \\
 1.107 & -0.117 & -1.512 & -2.479 & -0.774& -1.534 \\
\end{bmatrix}
\end{equation}
The Hamiltonian functions are given by (\ref{hfunction}). $h_1$ represents
a pure shear transformation and $h_2$ a pure rotation in phase space.

In a real physical process, the three systems (source, target and machine) 
will have their own Hamiltonian evolution. 
However, if we choose $\tau$ to be small (compared to any time scale present in the source, target and machine) we can
ensure that the cloning Hamiltonians $h_1$ and $h_2$ dominate over  the other terms. We can essentially assume that the 
evolution of the systems is ``frozen'' while cloning takes place.

\begin{figure}
\includegraphics[width=1\linewidth]{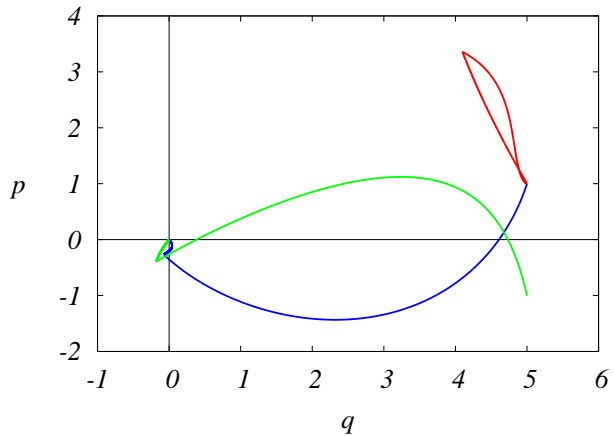}
\caption{Evolution of states (in phase space) for the 
Hamiltonians mentioned in \eqref{ham1} and \eqref{ham3} for the source (red), the target (blue) and the machine (green).}
\label{cl1}
\end{figure}

Figure \ref{cl1} shows the evolution of the source, the target, and the machine 
(in phase space) using the Hamiltonians in \eqref{ham1} and \eqref{ham3}, for a time, $t=2\tau$. 
From these figures we observe that, while the source returns to its original state after 
evolution i.e. $\begin{pmatrix} q \\ p \end{pmatrix}$, 
the target starts at the origin and goes to the 
state $\begin{pmatrix} q \\ p \end{pmatrix}$ making a perfect copy. 
The machine state also starts at the 
origin and it reaches $\begin{pmatrix} q \\ -p \end{pmatrix}$. 
If one considers the phase space area 
$A(D)=\int_D dq\wedge dp$ of a disk $D$, under cloning
this disk is duplicated in the target space and 
it would seem that the phase space area has doubled. 
However, the machine is also affected and the 
image of $D$ in the machine phase space is such
as to cancel out the duplication thereby \cite{aaron} preserving 
total phase space area. Thus we retain the original, and we have 
created a perfect clone and an anti-clone. (See Sec. $V$ for 
an explanation of the term ``anti-clone'').
The anti-clone contribution to the phase space area 
cancels the one from the clone, thus preserving the total phase space area.

Given three oscillators, one can, 
in principle create a perfect clone and anticlone along the 
lines described above by tuning the coupling strengths 
read off from (\ref{ham1},\ref{ham3}). This would be a minimal 
cloning machine, since the machine has the same phase space as the source.
In practice, this may need a large degree of control over the oscillators and their couplings.
We will describe below (sec. III) a more practical (albeit non-minimal) 
scheme for realising the cloning process in the laboratory using
non linear optics.  

\subsection{More General Cloning maps}
Fenyes' construction \cite{aaron} provides cloning maps for $\mathds{R}^{2n}$. 
This leaves open the question of cloning on other symplectic manifolds. A large class of 
symplectic manifolds appear in classical mechanics as cotangent bundles of configuration spaces. 
Some examples are the rigid rotor and the symmetric top with configuration space $S^2$ and $\mathds{R} P^3$ respectively.
Let us consider some illuminating examples.

Example 1: Starting with the cloning map above (\ref{clonmatrix}), notice that its entries are all integers.
We can therefore make an identification of $x$ with $x+1$, 
where $x$ is any of the coordinates of the $\mathds{R}^6$ phase space.
We thus compactify $\mathds{R}^6$ to $(T^2)^3$.
Since the identification is preserved under the 
action of $\phi$, we find a cloning map on $T^2$ viewed as a phase space. 
The symplectic structure on $\mathds{R}^2$ comes down to the quotient manifold 
$T^2$. Thus, minimal cloning machines on $T^2$ do exist.

Example 2: The cloning map (\ref{clonmatrix}) mixes the positions and momenta. The identifications made in the configuration space force similar identifications in the momenta.
In contrast, consider the following linear transformation 
on $\mathds{R}^3$, viewed as a `configuration space'. With $i=s,t,m$, and $\{q^i\}$, let
$$
q'^i=\Lambda^i_j q^j 
$$

\begin{equation}
\Lambda=\begin{bmatrix}[ccc]
 1 & -1& -1 \\ 
 1 & 0 & 1 \\ 
 1 & -1 & 0 
\label{clon2matrix}
\end{bmatrix}
\end{equation}
This linear transformation of the configuration space lifts to the 
cotangent space $T^* \mathds{R}^3$. The momenta transform under the inverse map
$A=\Lambda^{-1}$ as 
$$
p'_i=p_j A^j_i 
$$
where
\begin{equation}
A=\begin{bmatrix}[ccc]
 1 & 1& -1 \\
 1 & 1 & -2 \\
 -1 & 0 & 1
\label{clon3matrix}
\end{bmatrix}
\end{equation}
which clones the momenta. Thus the lift $\phi_2$ of the map $\Lambda$ 
to $T^* \mathds{R}^6$ is a symplectic, cloning map on the phase space. We now compactify the configuration space (using the earlier identification) and
get a cloning map on
$T^* S^1$. In fact, we could use coordinates $u^i=\exp{i q^i/(2\pi)}$ and turn the additive structure of the vector space 
of the original cloning map to a multiplicative structure of a group.
This suggests the next example.

Example 3: Let $Q$ be a group manifold, for example $S^3$ which can be identified with  the group $SU(2)$.
Consider the map $\sigma$
\begin{eqnarray}
u'^s&=&u^s (u^m)^{-1} (u^t)^{-1}\\
u'^t&=&u^s u^m\\
u'^m&=&u^s (u^t)^{-1}
\label{groupmap}
\end{eqnarray}
of the space $Q\times Q\times Q$ on to itself. The map is smoothly invertible,
$\sigma^{-1}$ being explicitly given by
\begin{eqnarray}
u^s&=&u'^s (u'^m)^{-1} (u'^t)\\
u^t&=&(u'^m)^{-1} u'^s (u'^m)^{-1}  u'^t\\
u^m&=&(u'^t)^{-1} u'^m (u'^s)^{-1}  u'^t
\label{groupmapinverse}
\end{eqnarray}
and so its lift to $T^{*}(Q\times Q\times Q)$ is automatically a symplectomorphism.
It follows that $\phi_3$ is a symplectic map
and it is easily checked that  $\phi_3$ clones on $T^*Q$, when the 
machine and target are initially in the identity state, with zero momentum.

\section{Proposed Experiment} 

As  mentioned earlier, classical cloning  can be realised by using techniques of non-linear optics\cite{nopbook}. In particular, one can use 
a four-wave mixing process like the Kerr effect to generate a clone and an anticlone (See Sec. V for an explanation of the term `anticlone'.), leaving the original unaltered. Each mode of the electromagnetic field  (characterised by a wave vector and polarisation) of the electromagnetic field is an oscillator with a definite frequency.
The phase space of an oscillator is $(q,p)$ and the complex number $q+ip$ can be thought of as a complex phase space coordinate.
Electromagnetic waves have negligible interaction with each other. The nonlinear crystal generates \cite{nopbook} an interaction between the electromagnetic modes.
This interaction takes place in the short time that the light spends in the non-linear crystal. We need only consider 
the participating beams before and after the interaction.

The ``machine'' here is considerably more complex than in the minimal cloning models of section II, since it includes  two pump beams
and a nonlinear material apart from the anticlone. 
We are concerned here with 
a third order nonlinear optics effect: the polarisation response of the material is cubic in the incident electric field: 
${\vec{P}}=\epsilon_0\chi^3 \vec E (\vec{E}.\vec{E})$, where $\epsilon_0$ is the dielectric permittivity and $\chi^{(3)}$ 
the third order susceptibility. 
(It follows from symmetry arguments that the second order susceptibility $\chi^{(2)}$ vanishes for centrosymmetric materials.)
The cloning machine consists of a non-linear sample illuminated by 
two strong`` pump'' beams.  These serve to bring the response of the sample into the nonlinear
regime, acting rather like the bias voltage of a transistor. When this sample is further illuminated by a weak ``signal'' beam, we find 
that in addition to the signal beam (the source) the system generates two more beams, a clone beam and an anti-clone beam. We will describe the scheme
more fully below.

Let us note 
first that a mode of the electromagnetic field is characterised by a wave vector $\vec{k}$ and a polarisation vector. We will keep
the polarisation vector fixed along $\hat{z}$ in the discussion below. All our wave vectors will lie in the $x-y$ plane. 
A wave in the $\hat{k}$ direction 
can be described by the $z$ component of its electric field $E=\hat{z}.\vec{E}$
\begin{equation}
E(\vec{r},t) = A u(\vec r,t)+\overline{A} \overline{u}(\vec r,t)
\label{wave}
\end{equation}
where 
$u(\vec r,t) = \exp{i(\omega t -\vec{k}}. \vec{r})$ and $A$ is a complex number which describes the amplitude and phase of the beam.

Each mode of the field is an oscillator with frequency $\omega=\tilde{c} |\vec{k}|$, where $\tilde{c}=c/n(\omega)$ is the speed of light
in the medium (which is assumed isotropic). The phase space of the oscillator is described by the real and imaginary parts of $A$ (the two quadrature components of the wave), 
which are canonically conjugate to each other. The symplectic form can be written $d \overline{A}\wedge d A/(2i)$. 
This symplectic form is clearly reversed by the map $A\rightarrow\overline{A}$ taking $A$ to $\overline{A}$.

We have supposed the medium to be isotropic, so that $\chi^{(3)}$ and $n(\omega)$, the refractive index are scalar. 
We suppose all beams in the experiment to have the same frequency $\omega$. This has the practical advantage that it 
makes it easier to satisfy the phase matching conditions \cite{nopbook}. 
Let us consider three incident  waves represented as follows:
\begin{equation}
E_j(\vec{r},t) = A_j u_j(\vec r,t)+\overline{A}_j \overline{u}_j(\vec r,t)
\label{efield}
\end{equation} 
where $u_j(\vec r,t) = \exp{i(\omega t -\vec{k}_j}. \vec{r})$ with $j=1,2,3$ and $\omega={\tilde{c}}{\vec{k}_j}$.
Here we consider the beam $1$ to be the signal beam (corresponding to the source). 
The beams $2$ and $3$ are the pump beams and the emergent beams contain the clone and the anticlone.
Thus there are actually five beams involved: one signal ${\vec k}_1$, two pumps ${\vec k}_2,{\vec k}_3$, one anticlone $-{\vec k}_1$ and one clone ${\vec k}_3+{\vec \delta}$. 
To start with, the clone and anticlone
are in their ground states. They are excited to the desired states by the interaction with the pump and signal beams.
Third order nonlinear processes are based on the term $\epsilon_0 \chi^{(3)} E^3$ in the expression for 
the polarisation. We are interested in the beams emerging at frequency $\omega$.
Expanding the cubic term $E^3$, the relevant terms in the polarisation 
are of the following form:
\begin{eqnarray*}\label{cubic}
P_a&=& \big[\epsilon_0 \chi^{(3)}A_2 A_3\big] \overline{A}_1  \exp{i(\omega t -(\vec{k}_2+\vec{k}_3-\vec{k}_1). \vec{r})} \\
P_c&=&\big[\epsilon_0 \chi^{(3)}\overline{A}_2 A_3\big] A_1  \exp{i(\omega t -(\vec{k}_1+\vec{k}_3-\vec{k}_2). \vec{r})}. \\
\end{eqnarray*}
Below we drop the constant terms in square brackets. These just indicate an overall change of amplitude and can be set to $1$ by judicious choice
of pump power. 

We now have to choose the $\vec{k}_i$'s so as to satisfy the phase matching conditions\cite{nopbook}, in both these beams.
We choose the wave vectors such that $\vec{k}_2+\vec{k}_3=0$ and $\vec{k}_1=\vec{k}_2$. 
For this choice the two terms mentioned 
above reduce to 
\begin{equation}
\label{anticlone}
E_a\propto{\overline{A}_1 \exp{i(\omega t +\vec{k}_1.\vec{r}))}},  
\end{equation}
which corresponds to an anticlone and 
\begin{equation}\label{clone}
E_c\propto{A_1 \exp{i(\omega t -\vec{k}_3.\vec{r})}},
\end{equation}
which corresponds to a clone. 
These two emergent beams satisfy the phase matching conditions, 
since in each case $\omega=\tilde{c}|\vec{k}|$ holds.

In the above arrangement, the directions of all the beams are collinear, which makes it awkward in a laboratory situation. 
For experimental ease, one can slightly perturb the direction of the $k_1$ beam by setting 
$\vec{k_1}=\vec{k_2}+\vec{\delta}$, such that $\vec{\delta}.\vec{k}_2 = 0$. 
Then $|\vec{k_1}+\vec{\delta}| \approx |\vec{k_1}|$ to first order 
in $|\delta|$ and thus we still satisfy the phase matching conditions, albeit approximately.
With this new scheme, the anticlone beam emerges in the $\vec{k}_a=-\vec{k}_1$ direction, while the clone beam emerges in the 
$\vec{k}_c=\vec{k}_3+\vec{\delta}$ direction, while the original source beam continues in the $\vec{k}_1$ direction from the linear part of the response.
Regarding $\omega$ as a carrier frequency we can use an Acousto-optic modulator (AOM) to impress a modulation on the signal $A_1$ so that $A_1$ depends
on $t$ on a slow timescale compared to the inverse carrier frequency. This results in an output in the clone channel (in the direction $\vec k_c$) proportional to 
$A_1(t)$ and in the anticlone channel (in the direction $\vec k_a$) proportional to $\overline{A}_1(t)$.

\section{Corruption of Classical Cloning by thermal noise}

Till now we have assumed an ideal, noise free situation in which the state of a system is described by a point in phase space. 
To understand the classical cloning process in a more realistic context, 
we introduce thermal noise to the system 
and study how this affects the cloning process.
In the ideal case, we had taken the states of the source, the target and the machine to be 
Dirac delta functions in the phase space. 
We now replace the delta functions with functions of finite width which are statistical mechanical probability distributions.
For the sake of convenience, we consider the distributions to be Gaussian. 
We suppose the source to be a Gaussian peaked about $(q_0,p_0)$. 
The source and machine are chosen to be Gaussians peaked around the origin.

The initial state of the total system (source, target, machine) is
taken to be
\begin{equation}
P_{in}(x)={\cal{N}} \exp{\big[-(x-\mu)^T {\cal A} (x-\mu)\big]}
\label{initialstate}
\end{equation}
where $x=\{q_s,p_s,q_t,p_t,q_m,p_m\}$ is a six dimensional vector,
$\mu=\{q_0,p_0,0,0,0,0\}$ represents the means of the initial distributions and 
${\cal A}$ a $6\times6$ diagonal matrix with diagonal entries 
$\{\alpha_s,\alpha_s,\alpha_t,\alpha_t,\alpha_m,\alpha_m\}$ and here and below, ${\cal N}$ is a normalisation constant. 
Under the cloning map
$x\rightarrow \Lambda x$, the distribution of the total system changes as
$P_{fin}(x)= J^{-1} P_{in}(\Lambda^{-1}x)$
(where $J=Det \Lambda=1$ since $\Lambda$ is symplectic).
\begin{equation}
{\cal{N}} \exp{\big[-(\Lambda^{-1}x-\mu)^T {\cal A} (\Lambda^{-1}x-\mu)\big]}
\label{finalstate1}
\end{equation}
which can be rewritten as
\begin{equation}
P_{fin}(x)= {\cal{N}} \exp{\big[-(x-\Lambda\mu)^T {\cal B} (x-\Lambda\mu)\big]}
\label{finalstate2}
\end{equation}
where ${\cal B}=(\Lambda^{-1})^T {\cal A} \Lambda^{-1}$.
It is evident that the means of the distributions are successfully cloned
$\mu'=\Lambda\mu$. As we will see below, the variances are not
faithfully cloned, in keeping with the classical no cloning theorem \cite{plastino}.

We can find the marginal distribution of the  
source by integrating over the target and machine. 
To do this we write ${\cal B}$ in block form

\begin{equation}
{\cal B}=\begin{bmatrix}[c|c]
 a\! & \,\,\,\,\,c^t   \\ \hline
   &     \\
 c\! & \,\,\,\,\,b  \\ 
\label{blockform}
\end{bmatrix}
\end{equation}
where $a$ is a non-singular $2\times2$ matrix $b$ a non-singular $4\times 4$ matrix
and $c$ a rectangular $4\times 2$ matrix and $c^t$ its transpose. 
It is straightforward to compute the marginal for the source.

This yields for the source distribution after cloning:
\begin{equation}
P_{sf}(q_s,p_s)={\cal N}_s \exp{\big[-(x_s-\mu_s)^T {\cal C}_s (x_s-\mu_s)\big]}
\label{sourcefinal}
\end{equation}
where $x_s=\{q_s,p_s\}$ is a $2$ dimensional vector, and
\begin{equation}
{\cal C}_s=a-c^t b c
\end{equation}
a $2\times 2$ covariance matrix and $\mu_s=\{q_0,p_0\}$. 

As an example, let us consider the target and the machine to be in
a thermal state with temperature $T$ with an oscillator Hamiltonian.
In fact, let us set $k=m=1$ in
the oscillator Hamiltonian $H=\frac{p^2}{2m}+\frac{1}{2}kq^2$ so that the frequency is $1$. 
The target Hamiltonian is $H_t=(q_t^2+p_t^2)/2$ and the machine Hamiltonian is $H_m=(q_m^2+p_m^2)/2$.
The Gibbs state of the target and machine is 
\begin{equation}
\text{P}=\frac{1}{Z}e^{-\beta (H_t+H_m)},
\end{equation}
where $\beta=\frac{1}{k_BT}$ and $Z$ a normalisation. 
We also set $\alpha_s=1$ and since the state of the machine and target are thermal, we have $\alpha_t=\alpha_m=\alpha=\beta/2$

The explicit form of the covariance matrix ${\cal C}_s$ is 
\begin{equation}
{\cal C}_s=\frac{1}{\Delta_s(\alpha)}\begin{bmatrix}
 \alpha^2+6\alpha & -4 \alpha \\
                  &            \\
                  &            \\
 -4\alpha & \alpha^2+4 \alpha  \\ 
\label{sourcecovariance}
\end{bmatrix}
\end{equation}
where $\Delta_s(\alpha)=\alpha^2+10\alpha+8$.

A very similar calculation, marginalising over the source and the machine 
gives the target state as

\begin{equation}
P_{tf}(q_t,p_t)={\cal N}_t \exp{-\big[(x_t-\mu_s)^T {\cal C}_t (x_t-\mu_s)\big]}
\label{targetfinal}
\end{equation}
where $x_t=\{q_t,p_t\}$ is a $2$ dimensional vector, 
${\cal C}_t$ a $2\times 2$ covariance matrix and $\mu_t=\mu_s=\{q_0,p_0\}$. 
The explicit form
of the covariance matrix ${\cal C}_t$ is 

\begin{equation}
{\cal C}_t=\frac{1}{\Delta_t(\alpha)}\begin{bmatrix}
 \alpha^2+3\alpha & \alpha \\
                  &            \\
                  &            \\
 \alpha & \alpha^2+2 \alpha  \\ 
\label{targetcovariance}
\end{bmatrix}
\end{equation}
where $\Delta_t(\alpha)=\alpha^2+5\alpha+5$.

As the general formulae make clear, in the limit of zero temperature $(\beta\rightarrow\infty)$, $\alpha$ goes to infinity and
the covariance matrices of both the source and the target go to the initial distribution: the cloning is perfect. However,
at finite temperature, there is corruption of the source as well as the target. There is also a spurious correlation between
momentum and position introduced by the cloning process. Thus, 
the cloning is imperfect, as expected 
from the classical no-cloning
theorem for classical systems with statistical distributions. 

Similar conclusions emerge from our numerical analysis,
which also shows how presence of statistical mechanical noise affects the cloning process. 
When noise is introduced either in the machine or the target state, 
the original gets corrupted and the copy (which is distinct from the corrupted original) 
is not perfect. For illustration, we describe only the case where 
the source and the target are delta functions. (Initially, $\alpha_s=\alpha_t$ are both infinite.) 
That is,  only the machine is noisy with an initial $\alpha_m=365$.
The means of the initial state to be copied are $\mu_p=8$ and $\mu_q=5$.

\begin{figure*}
\centering
\subfloat[Source with the means $(8,5)$, the standard deviations 
$(0.037,0.052)$, and the Correlation $0.708$.\label{sfig:testa1}]{%
 \includegraphics[height=6cm,width=.49\linewidth]{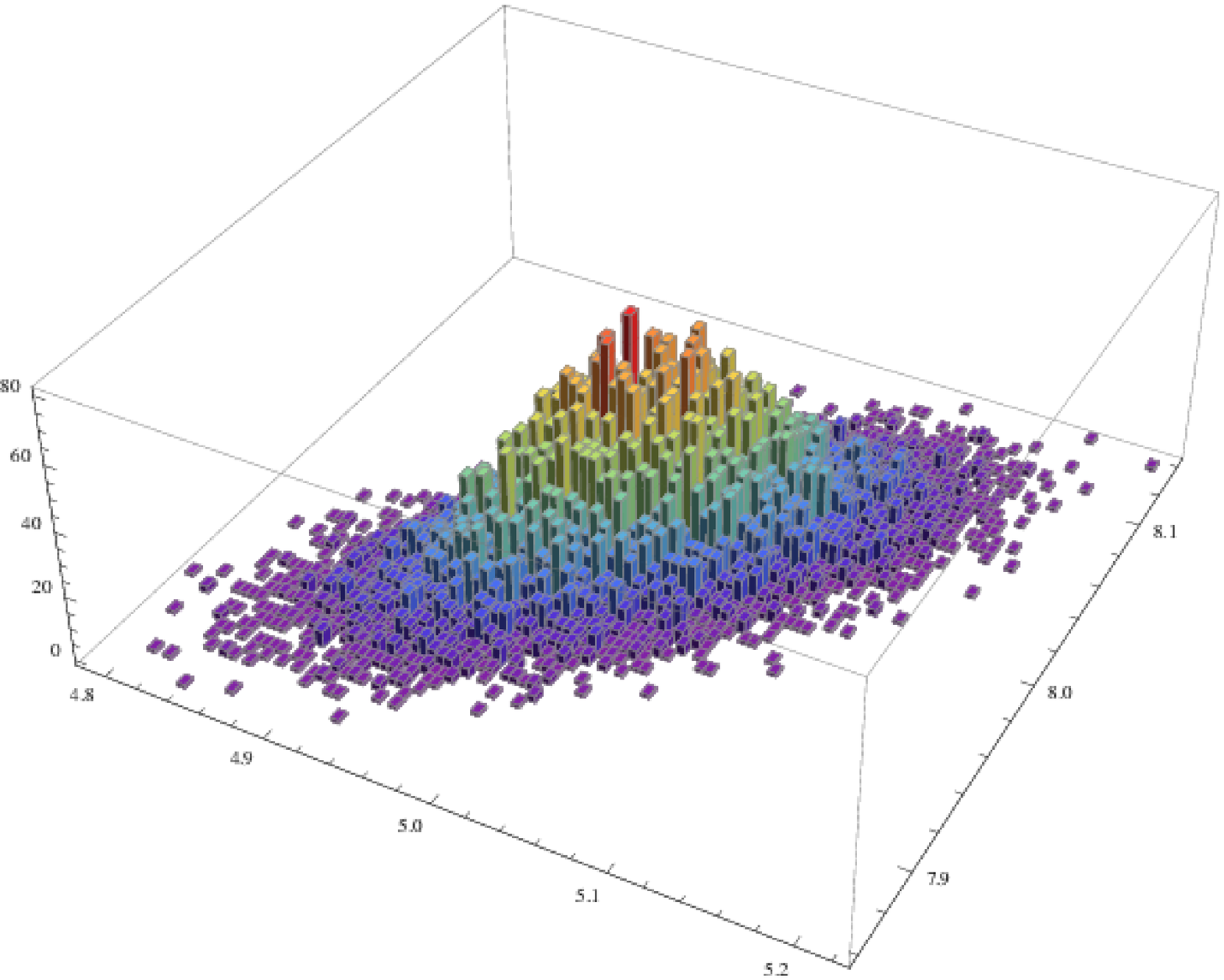}%
}\hfill
\subfloat[Target with the means $(8,5)$, the standard 
deviations $(0.03696,0.0371)$, and the Correlation $0.0017$.\label{sfig:testa2}]{%
  \includegraphics[height=6cm,width=.49\linewidth]{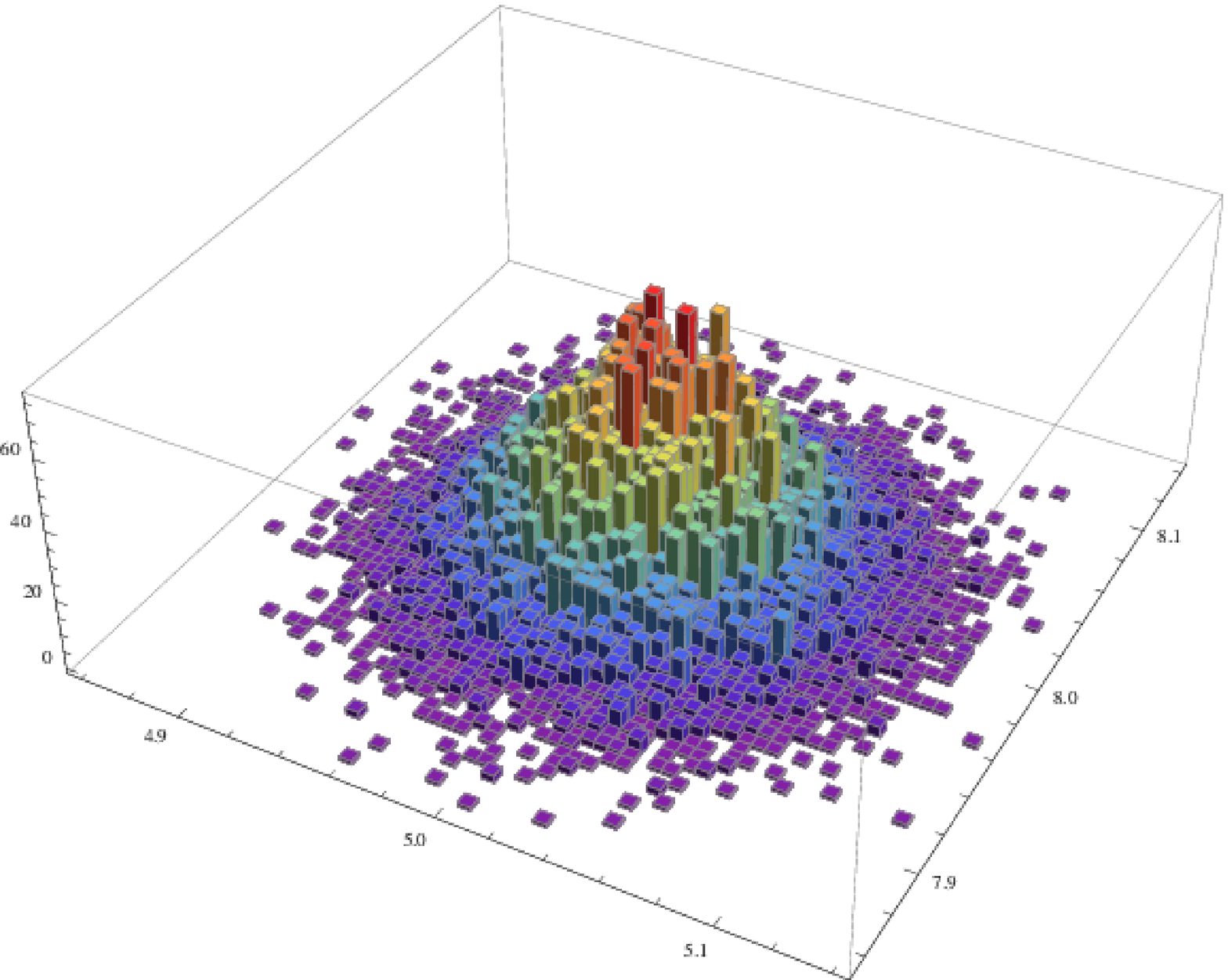}%
}\hfill
\subfloat[Machine with the means $(8,-5)$, 
the standard deviations $(0.052,0.083)$, and 
the Correlation $-0.32$.\label{sfig:testa3}]{%
  \includegraphics[height=6cm,width=.49\linewidth]{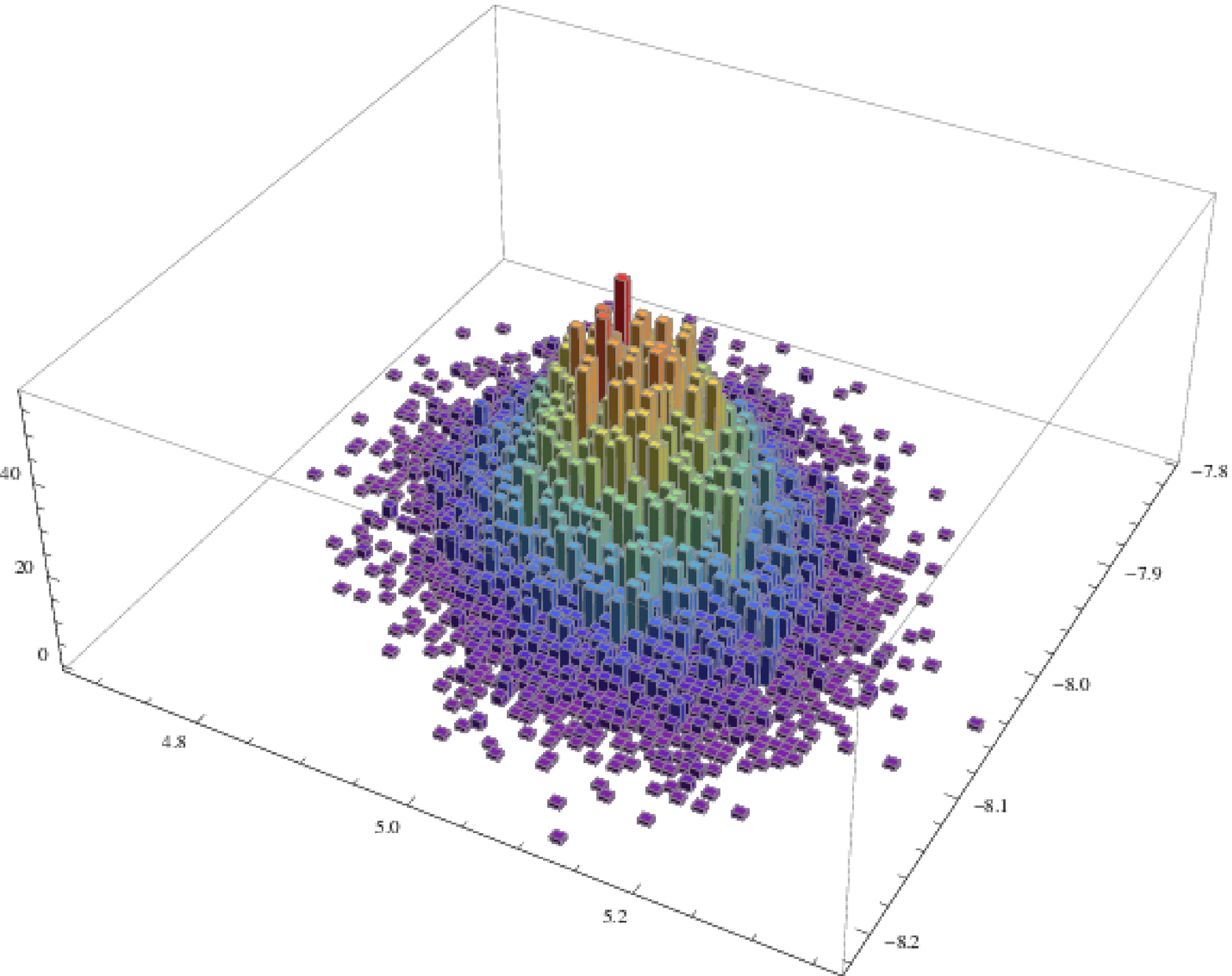}%
}
\caption{Histograms showing the distributions of the states in 
phase space after cloning, when noise is present only in the machine.}
\label{cr1}
\end{figure*}

Figure \ref{cr1} shows the distributions of the 
states after cloning when the machine is initially 
at an inverse temperature $1/(k_BT)= 730$.
This temperature is chosen so that the corruption is small but visible. 
Here it can be observed that all the three distributions are different from the original and from one another. 
These are Gaussian distributions and the most general form can be written as

\begin{equation}
\text{P}(q,p)=\frac{e^{ -\left(\frac{\left(q-\mu _q\right){}^2}{2 \sigma_q^2(1-\rho ^2)}-\frac{\rho  \left(p-\mu _p\right) \left(q-\mu _q\right)}{\sigma _p \sigma _q(1-\rho ^2)}+\frac{\left(p-\mu _p
\right){}^2}{2 \sigma _p^2(1-\rho ^2)}\right)}}{2 \pi  \sqrt{1-\rho ^2} \sigma _q \sigma _p},
\end{equation}
where, $\mu_q$, $\mu_p$ are the means and $\sigma _q$, $\sigma _p$ 
are the standard deviations pertaining to $q$ and $p$ and $\rho$ is the 
correlation between $q$ and $p$.

One can notice that (Figure \ref{cr1}), 
while the clones are imperfect, the means of the cloned states and the machine state 
are $(\mu_q,\mu_p),(\mu_q,\mu_p)(\mu_q,-\mu_p)$ i.e, the 
means are exactly where the perfect copies 
would be if the temperatures were  $0K$.
At any finite temperature, the target (and machine) will have deviations
from the blank (and ready) state. Since the map $\phi$ clones perfectly
{\it only} when there are no deviations, it is clear that thermal noise
corrupts the cloning process.

\section{Conclusion}

In summary, we have presented a discussion of classical cloning and its subtleties. 
We have given a systematic method for generating all possible 
linear cloning maps for $\mathds{R}^2$  and illustrated this method with an explicit example. 
We have gone beyond earlier literature\cite{aaron,plastino} 
in constructing explicit Hamiltonians generating a cloning map. We then propose a realisable experiment
to demonstrate  a classical cloning process in the laboratory, using non-linear optics. 
We have studied the effect of statistical noise on the cloning process. 
An important off shoot from this work is a proof
that any phase space emerging from a group manifold $Q$ (as a cotangent bundle $T^{*}Q$) 
admits a cloning map, in fact with a minimal machine size.

It is not presently clear whether  {\it all} symplectic manifolds admit cloning maps. 
For example $S^2$ can be given a symplectic structure, with the area on the standard sphere being the symplectic two-form.
We leave it for future work to determine
whether general symplectic manifolds admit cloning maps and what size of machine would be required for this.
It was also curious to note that we needed two quadratic Hamiltonians to generate the map $\phi$. Is this always true for cloning maps?
The answer is no, as shown by the following counterexample. Consider the map $\mathds{R}^3$ to $\mathds{R}^3$ given in (\ref{clon2matrix}). This matrix can be considered an element of
$SL(3,\mathds{R})$. Its matrix log is real and in the Lie algebra of $SL(3,\mathds{R})$. Lifting this map to $\mathds{R}^6$, we have a cloning map
which is realised by a single Hamiltonian.

Let us also remark in passing that despite the use of the words ``phase space'', and ``Liouville equation'',  in \cite{plastino},
their proof has nothing to do with the phase space of classical mechanics, the Liouville equation or symplectic structure. In fact, the proof offered
in Refs \cite{plastino},\cite{plastino2} is more general and  
goes through when the state space is an arbitrary oriented manifold, which could even be odd dimensional.  This point is
clarified below in an appendix. 

How does the cloning of states differ between quantum and 
classical mechanics? Intuition would suggest that our inability to clone 
is an essentially quantum phenomenon. Classical states, viewed as 
points in phase space, 
can be measured to any desired accuracy and therefore reproduced;
unlike quantum states which are disturbed by measurement.
However, some expositions  (the current Wikipedia version\footnote{ 
https://en.wikipedia.org/wiki/No-cloningtheorem Revision 
at 03:45, 15 February 2019}‎ for example)
of the quantum no cloning theorem do not include 
a ``machine'' or any ancillary degrees of freedom. They seek to copy 
the source state by a Unitary transformation of the source-target system.
This is a misleading argument, since under the same conditions, 
classical cloning is also forbidden\cite{aaron} under Hamiltonian evolution. 
Duplication of an arbitrary classical 
state also implies duplication of the phase space area of an arbitrary loop $A(\gamma)$
($A(\gamma)\rightarrow 2 A(\gamma)$). 
The machine {\it is} needed to cancel the excess 
phase space area. As we have seen, the presence of the machine 
renders cloning possible. The machine must at least be as large as the 
system to the cloned, but could be larger. 
In contrast, even with a machine present, quantum cloning is impossible by Unitary transformations, 
in accord with our intuition.

It is interesting to note that a clone state is always accompanied by an
``anticlone'' state. The anticlone is the 
final state of the machine, 
which is a symplectically reversed version of the original
state. In classical mechanics, reversal of symplectic structure can be interpreted as  time reversal, for instance $(q,p)\rightarrow(q,-p)$. 
More generally Hamilton's equations $\frac{df}{dt}=\{H,f\}$ are invariant under time reversal only if the symplectic structure is reversed.
Just as symplectic maps are the classical analogue of quantum unitary transformations, antisymplectic ones are the classical analogue of
quantum antiunitaries, of which time reversal is a prime example.
In our proposed experiment, we have taken
care to ensure that the anticlone is also manifestly 
present in one of the emergent beams. From the optics point of view, this
is a ``phase conjugate beam'', which is an implementation of time reversal. In fact, our proposed experiment is modelled 
very closely on the setup used in phase conjugation. 
A discussion of anticlone states also appears in Ref.\cite{RevModPhys.77.1225}
which treats quantum cloning.

In our discussion of a ``state'' in classical mechanics, we first introduced a state as a point in phase space
or as a statistical distribution with delta function concentration at a phase point. More general states in the statistical mechanical
sense emerged from convex combinations of these  ``states''. It is illuminating to compare this situation with quantum mechanics,
where ``pure states'' are rays in Hilbert space, or equivalently one dimensional projections. Convex combinations of ``pure states''
yield all possible quantum states or density matrices. In quantum mechanics, the no-cloning theorem applies even to ``pure states''.
In the classical case, ``pure states'' can be cloned, but statistical mixtures of ``pure states'' cannot. This seems to be an essential difference
between the classical and quantum cases. 

Another point worth stressing is that both thermal and quantum fluctuations spoil our ability to clone. This can be seen operationally
in the proposed experiment. Any thermal noise occurring in the pump beams will automatically leave its mark in both the clone beam
and the source beam. A similar effect happens with quantum fluctuations. Zero point fluctuations in the electromagnetic field
will cause spontaneous emission in the emergent beams and so spoil the cloning process.
We expect our study to generate interest in experimentally testing these ideas.

\section{Acknowledgements} 
\label{}
We thank Paulina Hoyos-Restrepo, 
Hema Ramachandran and Kumar Shivam for discussions.

\section*{Appendix A: A Remark on the Classical No Cloning Theorem}

The classical no-cloning theorem \cite{plastino} is based on the invariance of the Kullback-Leibler divergence under diffeomorphisms. 
As we noted in the main text,  the authors of Ref.(\cite{plastino}) use the language of phase space and Liouville measure. Our purpose
here is to note that their result is not related to phase space or the Liouville measure but is true on arbitrary manifolds.

Let $M$ be an n-dimensional orientable manifold and $\tilde{P}_1$, $\tilde{P}_2$ probability densities on $M$. The K-L divergence is defined as 
\begin{equation}
D_{KL}(\tilde{P}_1,\tilde{P}_2)=\int_M \tilde{P}_1 g[\tilde{P}_1/\tilde{P}_2]
\label{KDdefinition}
\end{equation}
where $g(x)=\log{x}$. 
This definition is clearly invariant under diffeomorphisms of $M$. Consider a vector field $v$, the infinitesimal generator of an arbitrary  diffeomorphism on $M$. In local coordinates, 
\begin{equation}
\frac{dx^a}{dt}=v^a
\label{evol}
\end{equation}
Let us convert the density $\tilde{P}_1$ into an $n-$ form $\alpha_1$ 
on $M$ by multiplying by the Levi-Civita tensor density 
$\underaccent{\tilde}{\eta}_{a_1...a_n}$ of weight minus one.
$\alpha_1= \underaccent{\tilde}{\eta}_{a_1...a_n} \tilde{P}_1$.
The time rate of change of $D_{KL}$ under such a diffeo is given by the integral of the Lie derivative of the integrand 
$\int_M \mathcal{L}_v(\alpha_1 g)$, which gives
$\int_M d(i_v \alpha_1 g)=\int_{\partial M} i_v\alpha_1 g$ which is a boundary term and vanishes if the distributions fall off fast enough. This 
proves the key result on which the main theorem of \cite{plastino} is based for
{\it all} manifolds, quite independent of any phase space structure. By 
appropriate choice of $g$, our 
proof also includes the more general version of the argument \cite{plastino2}

\section*{Appendix B: Construction of the Linear Symplectic Cloning map by 
Gram-Schmidt}

In this section, we will discuss the procedure for 
the generation of a symplectic cloning map. The cloning map 
is described by $(x,b,r)\to(x,x,Fx)$ (where, $b$ and $r$ are at the origin). 



Let us choose $(e_1,g_1,e_2,g_2,e_3,g_3)$ as a basis on $\mathds{R}^6$.
The symplectic form $\Omega$ is defined as follows: 
{\begin{eqnarray} \label{sym}
\Omega(e_i,e_j)&=&0, \ \Omega(e_i,g_j)=\delta_{ij}, \nonumber \\
\Omega(g_i,g_j)&=&0, \ \Omega(g_i,e_j)=-\delta_{ij}.
\end{eqnarray}
Let us consider $\phi:\mathds{R}^6 \rightarrow \mathds{R}^6$, an unknown linear map. 
$E_j=\phi e_j$ and $G_j=\phi g_j$. We are trying to choose $E_j$ and $G_j$ so that $\phi$ becomes a cloning map.
Let us consider 
\begin{eqnarray}
E_1&=&e_1+e_2+e_3, \\
G_1&=&g_1+g_2-g_3.
\end{eqnarray}

We now use a symplectic version \cite{mcduff1998introduction} of the  
Gram-Schmidt Orthonormalization procedure and construct the 
following:
\begin{eqnarray*}
\hspace*{-0.5cm} e'_2&=&e_2-\Omega(e_2,G_1)E_1+\Omega(e_2,E_1)G_1=-e_1-e_3, \\ 
\hspace*{-0.5cm} e'_3&=&e_3-\Omega(e_3,G_1)E_1+\Omega(e_3,E_1)G_1=e_1+e_2+2e_3, \\ 
\hspace*{-0.5cm} g_2&=&g_2-\Omega(g_2,G_1)E_1+\Omega(g_2,E_1)G_1=-g_1+g_3, \\ 
\hspace*{-0.5cm} g'_3&=&g_3-\Omega(g_3,G_1)E_1+\Omega(g_3,E_1)G_1=2g_3-g_1-g_2.
\end{eqnarray*}

We now construct 
\begin{eqnarray}
E_2&=&-e'_2-2g'_2+g_3=e_1+e_3+g_1-g_2, \\
G_2&=&e'_3-3G'_2+g_3, \nonumber \\
&=&e_1+e_2+2e_3+2g_1-g_2-g_3.
\end{eqnarray}

We then construct
\begin{eqnarray} 
e''_3&=&e'_2-\Omega(e'_2,G_2)E_2+\Omega(e'_2,E_2)G_2 \nonumber \\
&=& -e_1-e_2-2e_3-g_1+g_3, \\ 
g''_3&=&g'_2-\Omega(g'_2,G_2)E_2+\Omega(g'_2,E_2)G_2, \nonumber \\
&=&e_1+e_3-g_2+g_3.
\end{eqnarray}

We continue further to generate
\begin{eqnarray}
E_3&=&-g''_3=-e_1-e_3+g_2-g_3, \\
G_3&=&-e''_3=e_1+e_2+2e_3+g_1-g_3.
\end{eqnarray}

Our cloning map is given by Eq. (\ref{clonmatrix}).

\end{document}